\documentclass[aps,floats,showpacs,epsf,twocolumn]{revtex4-1}
\DeclareMathAlphabet{\mathpzc}{OT1}{pzc}{m}{it} 
\usepackage{amsfonts, color}
\usepackage{graphics}
\usepackage{graphicx}
\begin{document}

\title{Magnetic catalysis and axionic charge-density-wave in Weyl semimetals}
\author{Bitan Roy}
\email{Corresponding author: broy@umd.edu}
\affiliation{Condensed Matter Theory Center, Department of Physics, University of Maryland, College Park, MD 20742, USA}

\author{Jay D. Sau}
\affiliation{Condensed Matter Theory Center, Department of Physics, University of Maryland, College Park, MD 20742, USA}

\date{\today}

\begin{abstract}
Three-dimensional Weyl and Dirac semimetals can support a chiral-symmetry-breaking, fully gapped, charge-density-wave order even for sufficiently weak repulsive electron-electron interactions, when placed in strong magnetic fields. In the former systems, due to the natural momentum space separation of Weyl nodes the ordered phase lacks the translational symmetry and represents an axionic phase of matter, while that in a Dirac semimetal (neglecting the Zeeman coupling) is only a trivial insulator. We present the scaling of this spectral gap for a wide
range of subcritical (weak) interactions as well as that of the diamagnetic susceptibility with the magnetic field. A similar mechanism for charge-density-wave ordering at weak coupling is shown to be operative in double and triple-Weyl semimetals, where the dispersion is linear (quadratic and cubic, respectively) for the z (planar) component(s) of the momentum. We here also address the competition between the charge-density-wave and a spin-density-wave orders, both of which breaks the chiral symmetry and leads to gapped spectrum, and show that at least in the weak coupling regime the former is energetically favored. The anomalous surface Hall conductivity, role of topological defects such as axion strings, existence of one-dimensional gapless dispersive modes along the core of such defects, and anomaly cancellation through the Callan-Harvey mechanism are discussed.
\end{abstract}


\maketitle

\vspace{10pt}

\section{Introduction}

Three dimensional Weyl semimetals (WSMs) represent topologically nontrivial gapless systems that support linearly dispersing quasiparticle excitations with opposite chiralities in the vicinity of two so-called Weyl points that are separated in the momentum space \cite{volovik}. If Weyl fermions choose to reside at the same point in the Brillouin zone (BZ), which can occur at the transition point between storng $Z_2$ topological and trivial band insulators \cite{TI-review-1, TI-review-2, hassan-cava, ando, hassan-neupane, armitage, TPT-BiT}, the configuration is dubbed as Dirac semimetal (DSM). Due to momentum space separation of Weyl nodes, the time reversal and/or the inversion (parity) symmetry is broken in WSMs, which can then lead to peculiar electrodynamic responses, such as chiral-magnetic effect and anomalous Hall conductivity \cite{nielsen-ninomiya, burkov-balents, ingran, Zuynin-wu-burkov, aji, grushin, dtson, goswami-tewari, franz-vazifeh, sid-transport}.

Similar to monolayer graphene, three-dimensional WSMs or DSMs are also extremely robust against weak electron-electron interactions, due to the vanishing density of states [$D(E) \sim E^2$] near the apex of conical dispersions \cite{herbut-juricic-roy}. Nevertheless, if interactions are sufficiently strong, they can undergo phase transitions and enter into fully gapped massive phases \cite{aji-cdw, nandkishore-macienko}. However, the requisite interaction strength for such instabilities may be too high to realize any ordering in the pristine system. However, the application of strong magnetic fields can trigger the ordering tendencies even for weak interactions.

Placed in a magnetic field ($B$), the linear dispersion in WSM or DSM quenches into a set of Landau levels (LLs), and in particular the zeroth LL (ZLL) for the left and right chiral fermions are composed of non-degenerate and spin-polarized one-dimensional dispersive modes with energies $\pm v k_z$, respectively, where $v$ is the quasiparticle Fermi velocity. For the sake of simplicity, we here assume the spectrum to be isotropic. Therefore, weak enough electron-electron interaction can \emph{hybridize} the one-dimensional chiral ZLLs and develop a chiral-symmetry-breaking (CSB) spectral gap at the Weyl points \cite{catalysis-3D, catalysis-3D-PRL}. Similar instability may also occur within the ZLL of three dimensional non-relativistic Fermi liquids, when placed in strong magnetic fields \cite{yakovenko}. Due to the momentum space separation of Weyl nodes, the CSB mass breaks the translational symmetry and represents a \emph{charge-density-wave} (CDW) order~\cite{SCzhang-weylaxion, ingran}. The CDW order in WSMs stands as an example of \emph{axionic} state of matter that supports dynamic magneto-electric effect, captured by the $\bf{E} \cdot \bf{B}$ term and its coefficient is tied with the separation of Weyl points \cite{SCzhang-weylaxion}. The accompanying massless Goldstone mode or the \emph{sliding} mode in CDW phase is known as \emph{axion}. On the other hand, in DSMs the CSB order corresponds to a trivial insulator (neglecting the Zeeman coupling) since the Dirac points reside at the same point of the BZ.

We also address the competition between the axionic CDW and a spin-density-wave (SDW) orders. Both of them can led to a spectral gap within the ZLL. However, we show that while the CDW order pushes down all filled LLs (placed below the chemical potential), the SDW order causes spin-splitting of the filled LLs. Thus we believe (at least for sufficiently weak interactions) that CDW order is energetically favored over SDW.

At the quantum critical point between the topological and the normal insulators (in class AII), the Zeeman coupling ($\tilde{g}$) in a trivial DSM can give rise to left and right chiral fermions at isolated points in the BZ. Recent time has also witnessed the discovery of topological DSMs (two copies of superimposed WSMs protected by time-reversal, inversion and four-fold rotational symmetries~\cite{furusaki}) in Cd$_3$As$_2$ \cite{weylexperiment1}, Na$_3$Bi \cite{weylexperiment2}. In the presence of magnetic fields, the Zeeman coupling can separate the Weyl nodes in trivial and topological DSMs and support a WSM \cite{weyl-dirac-comment}. Various other proposals for realizing WSMs in condensed matter systems include pyrochlore iridates with antiferromagnet ordering \cite{vishwanath-iridate}, multilayer configuration of topological and normal insulators \cite{burkov-balents, Zuynin-wu-burkov}, and magnetically doped topological insulators \cite{magneticTI-1,magneticTI-2} etc. But, their experimental realization remains elusive so far. Rather in recent past material realization of WSMs has been confirmed in inversion-asymmetric TaAs~\cite{taas-1, tasas-2, taas-3}, NbAs~\cite{nbas-1}, TaP~\cite{tap-1}, and time-reversal-symmetry breaking YbMnBi$_2$~\cite{borisenko}, Sr$_{1-y}$MnSb$_2$~\cite{chiorescu}. Therefore, proposed many-body axonic ground state can possibly be observed in these materials in near future when these systems are placed in strong magnetic fields and we present the scaling behavior of the such order for a wide range of subcritical interaction with the magnetic field. Formation of a spectral gap at Weyl points can lead to measurable consequences in various physical quantities and here we address its impact on diamagnetic susceptibility (DMS). Since WSMs lives at the \emph{upper critical dimensions} ($d_{up}=3$), the scaling of the mass gap and DMS display \emph{logarithmic} corrections as $B \to 0$.

Our proposed mechanism for the generation of the axionic CDW order at weak coupling in the presence of strong magnetic field remains operative among the other members of the Weyl family, such as double-WSM and triple-WSM. Respectively these two systems support two- and three-fold degenerate, spin-polarized chiral-ZLL. Thus sufficiently weak interaction can hybridize them and give rise to a CDW order, and we expect that such broken symmetry phase can be realized in various double-WSM, such as HgCr$_2$Se$_4$~\cite{Fang-HgCrSe, bernevig, nagaosa}, SrSi$_2$~\cite{srsi2}.

In this work we address the correction to anomalous transport properties in various members of Weyl family in the presence of axionic CDW order. In the uniform phase the axionic CDW order gives rise to anomalous charge transport on the surface, which receive contribution from the one-dimensional chiral surface states as well as from bulk scattered stated, the gapped ZLLs. In the ordered phase the axionic CDW can also accommodate topological defects, such as line-vortex, also known as the \emph{axion string}. We here show that such topological defects hosts gapless one-dimensional dispersive mode in its core. Such gapless mode carry dissipationless current, which in turn is supplied radially from the bulk, according to the Callan-Harvey mechanism.

We now promote the organization principle for rest of the paper. In the next section, we derive the LL spectrum in WSMs, discuss the possibility of realizing various broken-symmetry phases and the competition between the CDW and the SDW orders at weak coupling. Sec.~III is devoted to address the change renormalization and diamagnetic susceptibility in the presence of a spontaneously generated spectral gap at the Weyl nodes. In Sec.~IV, we analyze the scaling behavior of the spectral gap and compare our results with a recent experiment. Generalization of magnetic catalysis mechanism for other members of the Weyl family (such as double- and triple-WSM) is discussed in Sec.~V. Discussions on chiral anomaly, anomalous transport, role of topological defects, Callan-Hervey mechanism in the axionic CDW phase are presented in Sec.~VI. We summarize our findings in Sec.~VII. The derivation of the gap equation in the presence of magnetic fields is relegated to the Appendix.

\section{Landau levels and magnetic catalysis}

We begin the discussion with the computation of the LL spectrum in WSMs. Let us define a four-component spinor $\Psi^\top(\vec{k})=[\Psi_{L}(\vec{k}), \Psi_{R}(\vec{k})]$, where $\Psi_{X}(\vec{k})$ are two component spinors, organized as $ \Psi_{X}^\top(\vec{k})=[ \Psi_{X, \uparrow} (\pm \vec{Q}+\vec{k}), \Psi_{X,\downarrow}(\pm \vec{Q}+\vec{k})]$ for $X=L,R$. Weyl nodes are located at $\pm \vec{Q}$, where nondegenerate, linearly dispersing left (L) and right (R) chiral bands cross zero-energy, respectively, and $\uparrow, \downarrow$ are the Kramers partners or two spin projections. For the sake of simplicity, we choose $\vec{Q}=Q \hat{z}$, whereas in pristine DSMs $|\vec{Q}|=0$. Response of these systems to electromagnetic fields ($\vec{A}$) is captured by the Hamiltonian 
\begin{equation}\label{DiracHgauge}
H[\vec{A},\vec{a}]= \sum_{j=1}^{3} i \gamma_0 \gamma_j (v \hat{k}_j -e A_j -\tilde{g} \; a_j \gamma_5 ),
\end{equation} 
in the low energy limit, where $e$ is the electronic charge. Mutually anticommuting $\gamma$ matrices are $\gamma_0=\tau_1 \otimes \sigma_0$, $\gamma_5=\tau_3 \otimes \sigma_0$, and $\gamma_j=\tau_2 \otimes \sigma_j$ for $j=1,2,3$. $\tau_0 (\sigma_0)$ and $\tau_j (\sigma_j)$ are respectively the two dimensional identity and standard Pauli matrices, operating on the chiral(spin) index. The external magnetic field $\vec{B}(=\vec{\nabla} \times \vec{A})=B \hat{z}$ is set to be along the $z$-direction. In Eq.~(\ref{DiracHgauge}), we have allowed the axial vector potential $a_3=B$ that supports Weyl points at $\pm \vec{Q}_Z$, where $\vec{Q}_Z =\tilde{g} B \hat{z}$ (say) \cite{weyl-dirac-comment, shovkovy-weyl}. For example, at the quantum critical point between topological and trivial insulators a massless Dirac fermions are realized at $\vec{k}=(0,0,0)$ point and application of magnetic field gives rise to Weyl points at $\pm \vec{Q}_Z$. The explicit dependence on the axial vector potential from Eq.~(\ref{DiracHgauge}) can, however, be eliminated by setting $\vec{Q}=\vec{Q}_Z$ in the spinor definition of $\Psi (\vec{k})$.

The orbital coupling of the uniform magnetic field supports a set of LLs at energies $\pm \sqrt{2 n B +v^2  k^2_z}$ [setting $a_j=0$ in Eq.~(\ref{DiracHgauge})] for $n=0,1,\cdots$, with degeneracy per unit area $\frac{2-\delta_{n,0}}{2 \pi l^2_B}$, where $l_B\sim \frac{1}{\sqrt{e B}}$ is magnetic length. The spin-polarized ZLL contains two branches of one-dimensional dispersive modes with energies $\pm v k_z$. The ZLLs are eigenstates of $\gamma_5$, the generator of \emph{chirality}, with eigenvalues $\pm 1$. In the continuum description $\gamma_5$ is also the generator of $U(1)$ translational symmetry in WSMs and $[H[\vec{A},\vec{a}],\gamma_5]=0$. An infinitesimal interaction can, therefore, hybridize the left and the right chiral ZLLs, and develop a mass gap at the Weyl points through a BCS-like mechanism (due to effective dimensional reduction of the system in the magnetic field). Such mechanism is known as \emph{magnetic catalysis} \cite{catalysis-3D, catalysis-3D-PRL}.

To gain insight into the nature of the CSB order, we consider a generic effective single-particle Hamiltonian
\begin{eqnarray}
H[\vec{m},\vec{n},Q_3] &=& H[\vec{A},0] + m_1 \gamma_0 + m_2 i \gamma_0 \gamma_5   \nonumber \\
&+& n_1 \gamma_3 + n_2 i \gamma_5 \gamma_3 + Q_{3} i\gamma_0 \gamma_3 \gamma_5,  \label{Haux}
\end{eqnarray}  
where $\vec{m}$ $=(m_1,m_2)=|\Delta|\left(\cos{\phi},\sin{\phi} \right)$ is the complex CDW order parameter, and $\phi$ is the $U(1)$ angle. Eigenvalues of $H[\vec{m},0, 0]$ are $\pm  \sqrt{2 n B + v^2 k^2_z + |\Delta|^2}$ for $n=0,1, \cdots$. Therefore, $\Delta$ introduces a spectral gap within the ZLL ($n=0$), and in addition also pushes all the filled LLs ($n \geq 1$) at negative energies further down. Hence, formation of Dirac masses within the ZLL is energetically quite favored.

In WSMs non-degenerate left and right chiral fermionic excitations live around $\pm \vec{Q}$, hence 
$$\Delta \sim \mbox{exp}(-2 i \; \vec{Q} \cdot \vec{r}) \: \: \langle c^\dagger_{\vec{Q}} \; c_{-\vec{Q}}\rangle,$$ 
represents a translational symmetry breaking CDW order, with periodicity $2 \vec{Q}$. Breaking of translational symmetry can be appreciated from the anti-commutation relation among two mass matrices appearing in Eq.~(\ref{Haux}) and the generator of translation $\{\gamma_0, \gamma_5 \}=\{i\gamma_0 \gamma_5, \gamma_5 \}=0$. Such translational symmetry breaking order can arise from the four-fermion interaction 
\begin{equation}
H^{CDW}_{int}= g \left[ \left( \Psi^\dagger \gamma_0 \Psi \right)^2 + \left( \Psi^\dagger i \gamma_0 \gamma_5 \Psi \right)^2\right],
\end{equation} 
which corresponds to the celebrated Nambu-Jona-Lasinio model for mass generation of relativistic fermions through spontaneous chiral symmetry breaking~\cite{NJL}.

Next we discuss the effect of the terms proportional to $n_1,n_2$ in Eq.~(\ref{Haux}). The spectrum in the effective Hamiltonian $H[0,\vec{n},0]$ is given by $\pm \sqrt{ \left( \sqrt{2nB}+ \sigma N \right)^2+v^2k^2_z}$ for $\sigma=\pm$, where $N=\sqrt{n^2_1+n^2_2}$. Degeneracy of the LLs is $\frac{1}{2 \pi l^2_B}$ for $n=0,1, \cdots$. The four-fermion interaction that supports such order is 
\begin{equation}
H^{SDW}_{int}= g' \left[ \left( \Psi^\dagger \gamma_3 \Psi \right)^2 + \left( \Psi^\dagger i \gamma_3 \gamma_5 \Psi \right)^2\right].
\end{equation} 
The fermionic bilnears $\langle \Psi^\dagger \gamma_3 \Psi \rangle=n_1$, $\langle \Psi^\dagger i \gamma_5 \gamma_3 \Psi \rangle=n_2$ represent two components of a translational symmetry breaking SDW order, which gaps out the ZLL ($n=0$), but spits the filled LLs, placed below the chemical potential. Therefore, the SDW order, although introduces a spectral gap within the ZLL, is expected to be energetically inferior to the CDW order, which besides gaping the ZLL out, also pushes down the filled LLs. Therefore, we strongly believe that the CDW is the natural ground state in WSMs, when they are placed in strong magnetic field, at least for weak interactions. Thus, from now on we only focus on the CDW order.

In Eq.~(\ref{Haux}) we also allowed a term ($\sim Q_{3}$) that can be dynamically generated by electronic interactions and renormalize the location of Weyl points~\cite{shovkovy-weyl}. Following the same procedure, described above, $ Q_{3}$ can be eliminated from Eq.~(\ref{Haux}) by taking $Q \hat{z} \to (Q + Q_{3})\hat{z}$ in the definition of spinor $\Psi(\vec{k})$. However, a finite $Q_{3}$ modifies the periodicity of CDW order to $ 2 (Q  + Q_{3})\hat{z}$~\cite{comment-periodicity}. The four-fermion interaction $g_5 \left( \Psi^\dagger i \gamma_0 \gamma_3 \gamma_5 \Psi \right)^2$ can, in priciple, renormalize the location of the Weyl nodes, where $Q_3 \sim g_5 (l_B) \langle \Psi^\dagger i \gamma_0 \gamma_3 \gamma_5 \Psi \rangle$, where $g_5 (l_B)$ is the strength of four-fermion interaction $g_5$ at the scale $l_B$.

Formation of the CDW order can also take place when the system is placed slightly away from the charge-neutrality point. A CDW order, with periodicity $2 |Q + \frac{|\mu|}{v}|$ develops in WSMs at finite chemical doping ($\mu$), at least when $|\mu|< \sqrt{2B}$. In Cd$_3$As$_2$, Na$_3$Bi the field induced (by Zeeman coupling) Weyl nodes are located at $(\pm Q_0 \pm Q_z)\hat{z}$~\cite{dai}. Hence, the periodicity of field induced CDW order in these topological DSMs is $2 |Q_z|$ (assuming the periodicity of the CDW order is unique). The periodicity of the CDW order can be measured by scanning tunneling microscope, for example. If we completely neglect the Zeeman coupling and set $Q_3=0$ in DSM, one enjoys the liberty of setting $\phi=0$ and $\Delta(=m_1)$ then represents a trivial Dirac mass. However, scaling of DMS ($\chi$) and mass gap ($\Delta$) is insensitive to the exact nature of the CDW ordering.

\section{Charge renormalization and Diamagnetic susceptibility}

\begin{figure}
\includegraphics[width=8.0cm,height=5.75cm]{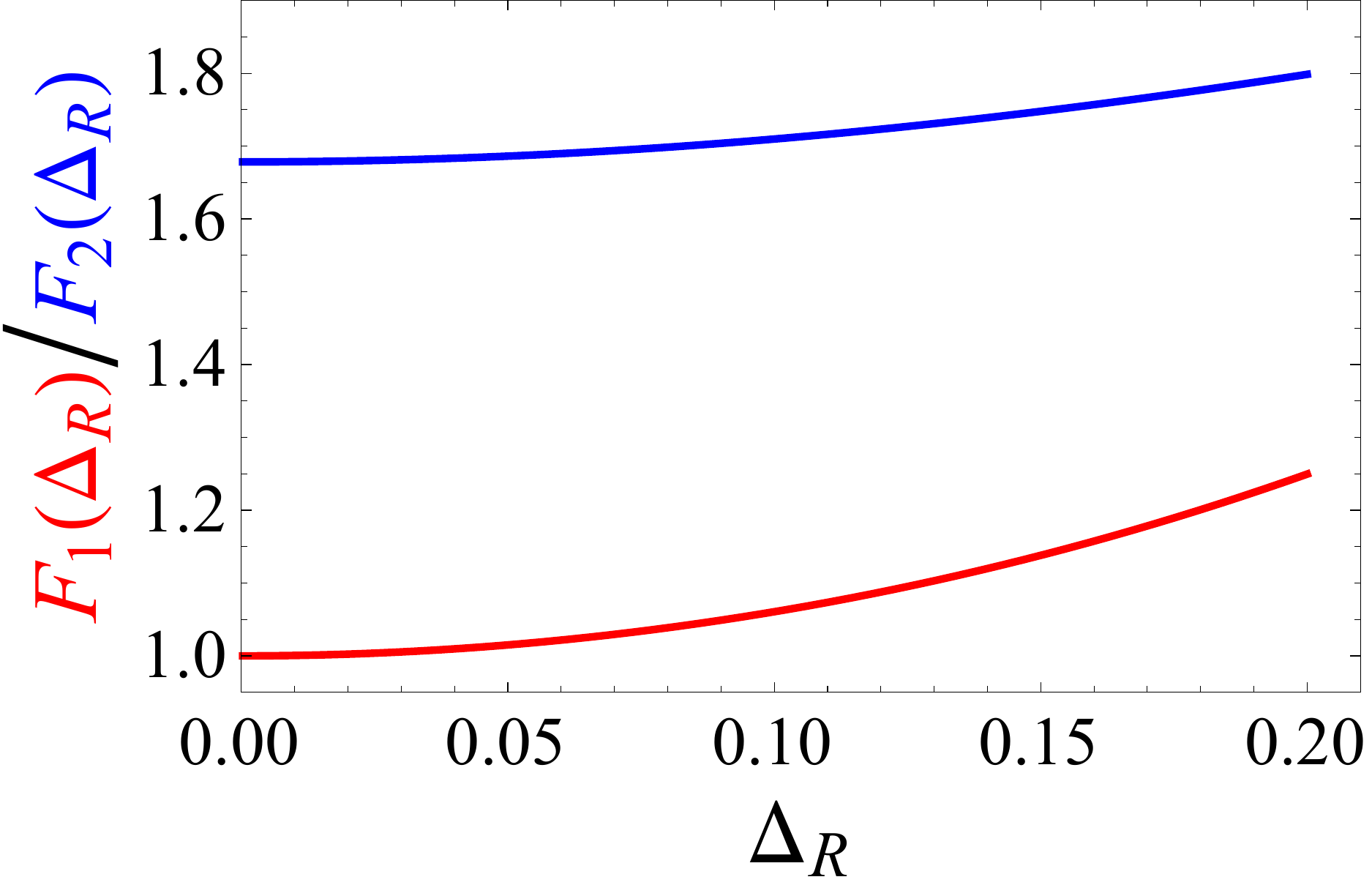}
\caption[] {(Color online) Scaling of two functions $F_1$ and $F_2$, appearing in the expression of DMS [see Eq.~(\ref{DMS})], with $\Delta_R$.}\label{fig-1}
\end{figure}

We now analyze the effects of mass generation near the Weyl points in the presence of magnetic fields on DMS and renormalization of electric charge and magnetic field. In magnetic fields, the free energy density ($F$) of the system scales as $F \sim \sqrt{e B}/l^3_B$, where $\sqrt{eB}$ is LL energy, and thus $F \sim (e B)^2$. Hence, a naive scaling argument indicates a \emph{constant} DMS $(\chi)$ in WSMs. However, free energy and DMS receive logarithmic corrections, since the system lives at upper critical dimension, which we capture pursuing a field theoretic approach \cite{abdassalam, goswami-chakravarty}. For simplicity, we consider a constant CSB mass ($\Delta$). The DMS then acquires contribution only from the higher LLs ($n \geq 1$) and the free energy is given by
\begin{widetext}
\begin{eqnarray}
F &=& \left( -\frac{1}{\pi l^2_B} \right) \times \int^{\infty}_{-\infty} \frac{d k_z}{2 \pi} \sum^{\infty}_{n=1} \sqrt{v^2 k^2_z+\Delta^2+n \frac{2 v^2}{l^2_B}}
=\left( -\frac{v}{\pi l^2_B} \right) \:  
\times \lim_{\epsilon \to 0}  \int^{\infty}_{-\infty} \frac{d k_z}{2 \pi} \sum^{\infty}_{n=1} \left[ \frac{2}{l^2_B \Lambda^2} \left[ \frac{k^2_z l^2_B}{2} +\frac{\Delta^2 l^2_B}{2 v^2} +n \right] \right]^{\frac{1}{2}-\frac{\epsilon}{2}} \nonumber \\
&=& \left( -\frac{v^2}{2\pi^2 l^4_B}\right) \times \bigg[ \frac{2 H\zeta (-1, 1+\Delta^2_R)}{\epsilon} 
+ \bigg\{ \log\left( \frac{\Lambda^2 l^2_B}{2}\right)
+ 0.386 \bigg\} \times H\zeta(-1,1+\Delta^2_R) +H\zeta'(-1,1+\Delta^2_R) \bigg] \nonumber \\
&=& \left( \frac{v^2 }{24 \pi^2 l^4_B} \right) \:\: \times \left( \frac{2}{\epsilon} \; F_1(\Delta_R)
+ \bigg[ \log\left( \frac{\Lambda^2 l^2_B}{2} \right) + 0.386 \bigg] \; F_1(\Delta_R) - F'_1 \left( \Delta_R \right) \right),
\end{eqnarray}
\end{widetext}
where $\Lambda$ is the ultraviolet (UV) cutoff for the conical dispersion in WSM, $\Delta_R= \frac{\Delta l_B}{\sqrt{2}v}$, and $H\zeta$ is the Hurwitz zeta function. The function 
\begin{equation}
F_1(x)=-12 H\zeta[-1,1+x^2],
\end{equation}
with $F_1(0)=1$ and its scaling is shown in Fig.~\ref{fig-1}. The term proportional to $ \frac{1}{\epsilon}$ can be identified as the logarithmically divergent piece in the free energy, which can be removed through the renormalization of electric charge $(e)$ and magnetic field $(B)$ according to  
\begin{eqnarray}
e^2_{ R} &=& e^2 \left[1-\frac{e^2 v}{12 \pi^2} \: F_1 (\Delta_R) \: \times \frac{1}{\epsilon} \right], \\
B^2_{ R} &=& B^2 \left[1+\frac{e^2 v}{12 \pi^2} \: F_1 (\Delta_R) \: \times \frac{1}{\epsilon} \right]
\end{eqnarray}    
where the quantities with subscript $R$ represent their renormalized values. The logarithmic correction in the free-energy is determined by the largest energy scale among $\sqrt{eB}$ and $\Delta$, which sets the infrared cut-off of the theory. For weak interactions $\sqrt{eB} \gg \Delta$, and the logarithmic correction is given by $\log(B_0/B)$, where $B_0 \sim \Lambda^2$ is the magnetic field associated with lattice spacing.

The finite part of free energy gives DMS 
\begin{eqnarray}\label{DMS}
\chi=\left( -\frac{e^2_R v}{24 \pi^2} \right) \:\: \left[ F_1 (\Delta_R) \log\left( \frac{B_0}{B}\right) + F_2 (\Delta_R)\right],
\end{eqnarray}   
where the function
\begin{equation}
F_2(x)=12 \big[ 0.31 H\zeta[-1,1+x^2] -H\zeta'(-1,1+x^2) \big].
\end{equation} 
Scaling of $F_2(x)$ is shown in Fig.~\ref{fig-1} and $F_2(0)=1.68$. Therefore, DMS in WSMs in addition to a constant value also manifests a \emph{logarithmic} enhancement as $B \to 0$, which can be measured in the small field limit. In addition, Eq.~(\ref{DMS}) suggests that DMS, besides the logarithmic correction, acquires non-trivial contributions due to the gap generation in magnetic field (see Fig.~1). We expect such corrections to DMS can be seen in experiments. 

\section{Scaling of spectral gap}

In the previous section, we assumed that the spectral gap at the Weyl point is insensitive to the strength of the magnetic field. However, the CSB mass displays nontrivial dependence on magnetic field and interaction strength, which we explore in this section. The qualitative behavior of DMS remains same, even when $\Delta=\Delta(B)$. 

\begin{figure}
\includegraphics[width=8.0cm,height=5.75cm]{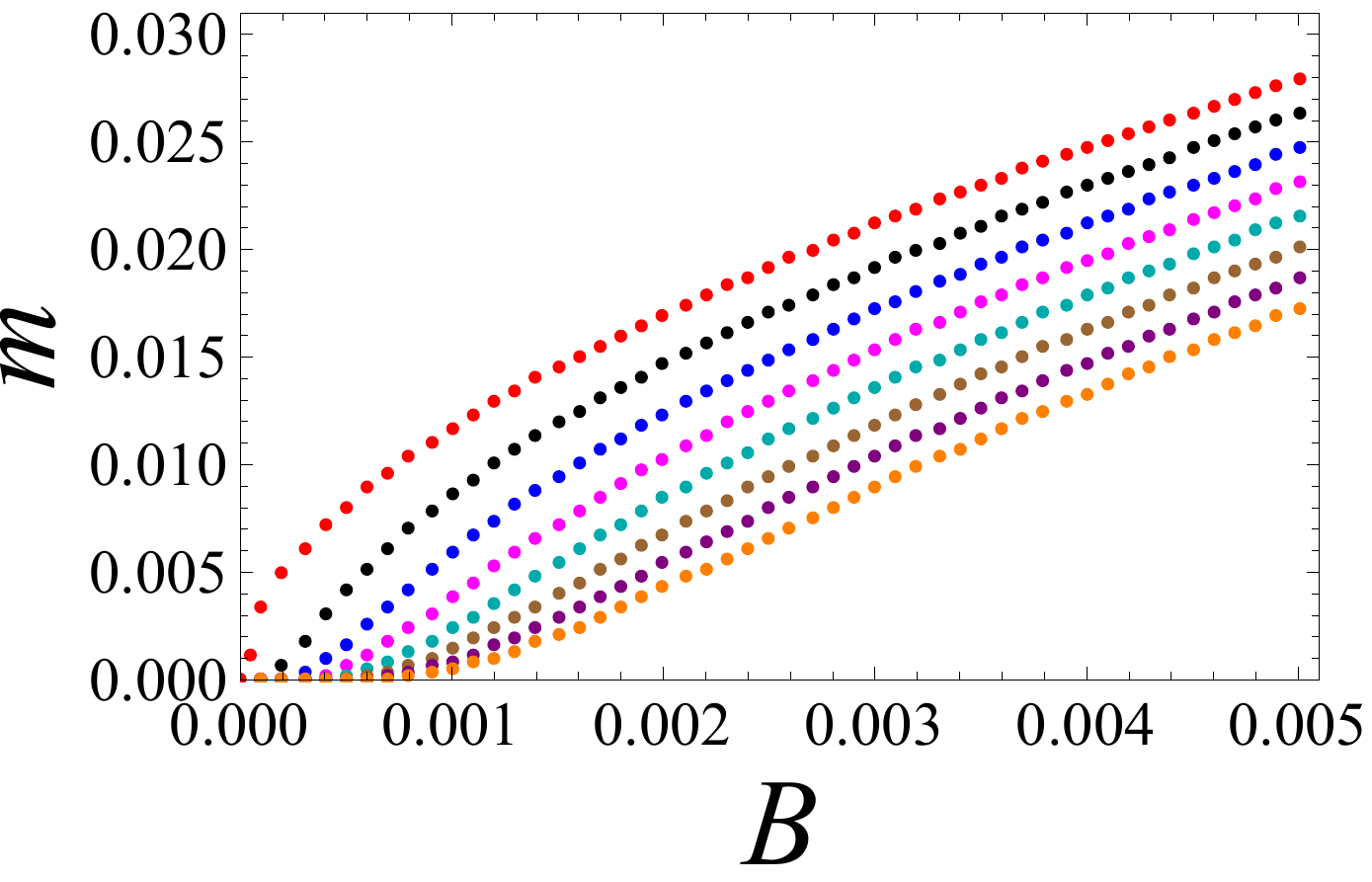}
\caption[] {(Color online) Scaling of CSB mass ($m$) with the magnetic field ($B$) for interaction strength, decreasing from top to bottom, parametrized by $\delta(=(g_c-g)/g g_c \Lambda^2)=0$, $0.01$, $0.02$, $0.03$, $0.04$, $0.05$, $0.06$, $0.07$. Here, $m$ is measured in units of $v \Lambda$, and $B$ in units of $B_0 \sim \Lambda^{2}$, magnetic field associated with the lattice spacing ($a \sim 1/\Lambda$).}\label{fig-2}
\end{figure}

The condensation energy in the presence of the CSB order is
\begin{equation}\label{freeen-gap}
E=\frac{\Delta^2}{4 g} -\int^{\infty}_{-\infty} \frac{d k_z}{2 \pi} \sum^{\infty}_{n=0}(2-\delta_{n,0}) 
\frac{\sqrt{2 n B +v^2 k^2_z + \Delta^2}}{2 \pi l^2_B}.
\end{equation}
We here consider only the short-range component ($g$) of Coulomb interaction that supports CSB order. Minimizing $E$ with respect to $\Delta$, we obtain the gap equation~\cite{herbut-roy-scaling}
\begin{eqnarray}
\frac{1}{g}=B \int^{\infty}_{\Lambda^{-2}} \frac{ds}{s} e^{-s \Delta^2} \; \coth(sB),
\end{eqnarray}
after completing the integral over $k_z$ and taking $2\pi^2/g \to 1/g$. A detailed derivation of the gap equation is avaliable in the Appendix. The right hand side of the above equation discerns an UV divergence as we take $\Lambda \to \infty$, which can be regularized by introducing a quantity $\delta=(g \Lambda^2)^{-1}-(g_c \Lambda^2)^{-1}$ that measures the deviation from the zero magnetic field critical strength of the interaction ($g_c$) for the CSB ordering, defined as $ g^{-1}_c= \int^\infty_0 dx K(x)/x^{2}$. The function $K(x)$ satisfies $K(x \to 0/\infty)= 0/1$, otherwise arbitrary. In terms of dimensionless gap $\frac{\Delta}{v \Lambda} \to m (\ll 1)$ and magnetic field $\frac{B}{\Lambda^{2}} \to B (\ll 1)$, the gap equation simplifies to 
\begin{equation} \label{gap-compact}
\delta+ I_1(m, B)+ I_2 (m, B)+ {\cal O}(m^6, B^4, y^{-4})=0, 
\end{equation}
where 
\begin{eqnarray}
 \frac{I_1 (m,B)}{2B} &=& a- \frac{b}{y^2}+\frac{c}{y^3} - \frac{y+1}{2 y} \left[\log(2B)- B+\frac{B^2}{6} \right], \nonumber \\
 \frac{I_2 (m,B)}{B} &=& \gamma_E + 2 \log(m) -m^2 + \frac{m^4}{4}, 
\end{eqnarray} 
and $y(=B/m^2) \gg 1$, for subcritical ($g<g_c$ or $\delta>0$) strength of the interaction. $\gamma_E$ is the Euler-Mascheroni constant, and $a=0.63$, $b=0.21$, $c=0.05$ (see Appendix). Numerical solutions of the above gap equations for a wide range of subcritical interaction are presented in Fig.~\ref{fig-2}. The mass gap acquires \emph{logarithmic} corrections as $B \to 0$, since the system lives at upper critical dimension ($d_u=3$).

So far we have considered only the short-range pieces of the Coulomb interaction, and neglected its long range tail since dielectric constant in semiconductors are typically very high ($\sim 10-30$). Weak long range interaction is a \emph{marginally} irrelevant perturbation in WSMs and gives rise to \emph{logarithmic} correction to Fermi velocity ($v$) \cite{goswami-chakravarty, nagaosa-2, migdal} according to $v \approx v_0 \left[1 + \alpha \log \left( B/B_0\right)\right]$ in magnetic fields, where $v_0$ is bare Fermi velocity and $\alpha$ is the fine structure constant. Therefore, mass gap ($m$) also acquires additional logarithmic correction, since $m$ is here measured in units of $v \Lambda$ \cite{herbut-roy-scaling}. The DMS ($\chi$), quoted in Eq.~(\ref{DMS}), also receive additional logarithmic correction from the long range tail of the Coulomb interaction, as $\chi$ is expressed as function of $\Delta_R=\Delta l_B/(\sqrt{2}v)$.

\begin{figure}
\includegraphics[width=8.0cm,height=5.75cm]{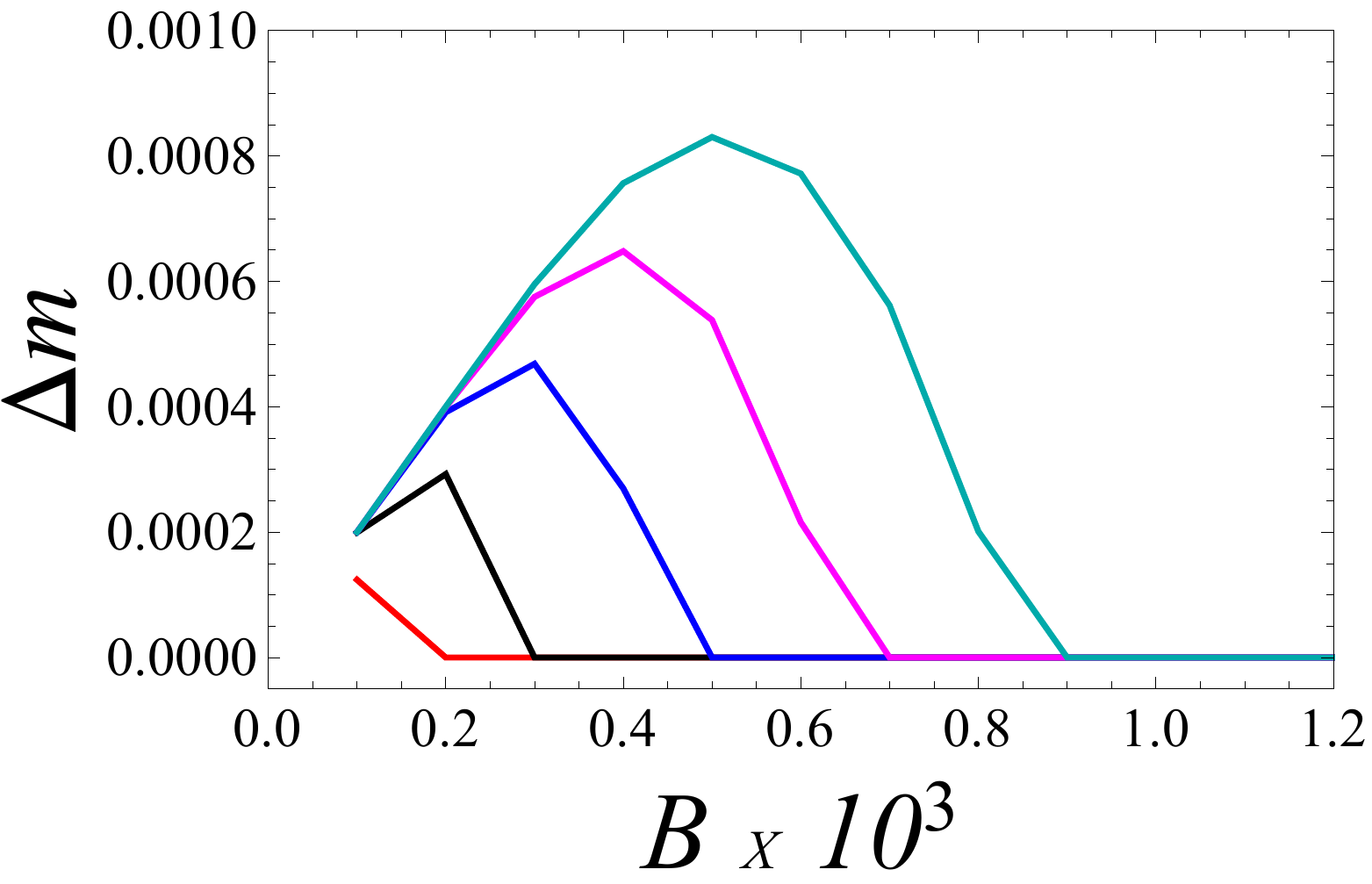}
\caption[] {(Color online) Scaling of the difference between the total gap $m_t(=m_b+m)$ and the interaction driven gap ($m$), $\Delta m=m_t-m$ as a function of dimensionless magnetic fields $B=B/\Lambda^2$, for $\delta=0.01$ (red), $0.02$ (black), $0.03$ (blue), $0.04$ (magenta), $0.05$ (cyan).}\label{fig-3}
\end{figure}

In a recent experiment, LL quantization has been observed in Cd$_3$As$_2$ \cite{cdas-LL}. However, the crystal has been cleaved along a low symmetry axis $(112)$, and consequently the underlying $C_4$ symmetry, protecting the gapless semimetallic phase~\cite{furusaki}, is lost. Therefore, even the non-interacting Hamiltonian gives rise to a gap in the spectrum, which possibly scales as $m_{b}=m_0 B$, and $m_0$ is chosen such that at $B=12$T, $m_b$ produces a non-interacting gap $1.4$ meV, in qualitative agreement with Ref.~\onlinecite{cdas-LL}. Performing self-consistent calculation of total gap $m_t=m+m_{b}$ we find that for weak magnetic fields the non-interacting gap $m_b$ dominates over interaction driven mass gap ($m$), that overwhelms the former contribution at stronger fields. Such crossover takes place at stronger magnetic fields as the interaction gets weaker, as shown in Fig.~\ref{fig-3}.

Cd$_3$As$_2$, Na$_3$Bi cleaved along high-symmetry axis, so that underlying $C_4$ symmetry is preserved, provides the ideal situation to observe only interaction induced gap at Weyl points. Alternatively, one can also compare the scaling of this gap with magnetic field at different temperatures. At high temperature the gap is determined by its non-interacting component ($m_b$), which is expected to scale linearly with the magnetic field, whereas at sufficiently low temperatures interaction driven gap ($m$) can take over $m_b$. Notice the critical temperature for CDW transition is $T_c \sim \exp(-1/[g D(B)])$, where $D(B)$ is the density of states of the ZLL (BCS-scaling). Thus subtracting $B$-linear piece of the gap, obtained from its high temperature scaling, one can extract the scaling behavior of the interaction induced gap in magnetic fields at low temperatures. Therefore, CSB mechanism for insulation in WSMs can be identified from the temperature dependence of magnetic field induced gap \cite{cdas-LL} and also from the scaling of DMS.

\section{Magnetic catalysis in double- and triple-Weyl semimetals}

The magnetic catalysis for gap formation in WSMs at weak coupling remains operative for other members of the Weyl family, such as double- and triple-WSMs. Respectively, these two systems display quadratic and cubic dispersions in the $x-y$ plane and a linear dispersion along the $z$-direction in the vicinity of Weyl nodes at $\pm \vec{Q}$. For example, double-WSMs can be realized in HgCr$_2$Se$_4$ \cite{bernevig, Fang-HgCrSe, nagaosa} and SrSi$_2$\cite{srsi2}, although the material realization of triple-WSM remains illusive so far. The bulk topological invariants in double- and triple-WSMs are respectively twice and thrice that in WSM, and consequently the the one-dimensional chiral surface states carry additional two and three fold degeneracy.  

The low-energy Hamiltonian in double-WSMs, placed in a magnetic field $\vec{B}=B \hat{z}$, reads as 
\begin{equation}
H_2[\vec{A}]=i \gamma_0 \left[ \gamma_1 \; \left( \frac{\pi^2_x -\pi^2_y}{2 m^\ast} \right) + \gamma_2 \; \left( \frac{2 \pi_x \pi_y}{2 m^\ast} \right) + \gamma_3 v k_z\right],
\end{equation}     
where $\pi_j =\left( -i \partial_j-A_j \right)$ and $m^\ast$ is the effective mass of the parabolic dispersion in $x-y$ plane. The spectrum of LLs goes as $\pm \sqrt{n (n-1) \omega^2_c +v^2 k^2_z}$ for $n=0,1,2, \cdots $, where $\omega_c$ is cyclotron frequency. Therefore, double-WSMs also host one dimensional spin-polarized chiral ZLLs, which, however, carry an extra two-fold \emph{orbital degeneracy} (for $n=0,1$)~\cite{bitan-BLG}. Hence, weak repulsive interactions can hybridize the chiral ZLLs and a CSB gap ($\Delta$) opens up at the double-Weyl points. The ZLL are then placed at $\pm \sqrt{v^2 k^2_z+\Delta^2}$. 

The effective Hamiltonian for tripple-WSM reads as 
\begin{equation}
H_3[\vec{A}]=i \gamma_0 \left[ \gamma_1 \; \frac{\pi^3_+ + \pi^3_-}{\Gamma} + \gamma_2 \; \frac{\pi^3_+ - \pi^3_-}{\Gamma} + \gamma_3 v k_z \right],
\end{equation}    
in a magnetic field $\vec{B}=B \hat{z}$, where the parameter $\Gamma$ controls the curvature of the cubic dispersion, and $\pi_\pm=\pi_x \pm i \pi_y$. The spectrum of LLs is given by $\pm  \sqrt{ \frac{l^{-6}_B}{\Gamma^2 } \: n(n-1) (n-2) + v^2 k^2_z}$. Hence, the chiral-ZLL in triple-WSM carries a three-fold orbital degeneracy (for $n=0,1,2$). Thus, sufficiently weak repulsive interactions can hybridize the chiral-ZLL and open a spectral gap at the triple-Weyl points. The additional two and three fold degeneracy of the ZLL, respectively in double- and triple-WSMs stem from the bulk topological invariant of these two systems. Due to such additional degeneracy of the one-dimensional chiral-ZLL, we expect that the transition temperature for the axionic CDW order in these materials to be higher than that in WSMs (neglecting the effect of disorder).     

Note the ZLLs in double- and triple-WSMs can also be gaped out by the SDW order. However, in these systems as well, the SDW order splits the filled LLs. Hence, we expect that in double- and triple-WSMs the CDW order is energetically favored over the SDW order.

\section{Axionic density-wave, chiral anomaly and topological defects}

Momentum space separation of Weyl nodes gives rise to a translational symmetry breaking CDW order at weak coupling in the presence of magnetic fields, which enters Eq.~(\ref{Haux}) as a complex mass $\Delta=m_1 + i m_2$ and the $U(1)$ angle ($\phi$) between $m_1$ and $m_2$ is a dynamic variable. Thus, CDW order in Weyl semimetals represents an axionic state of matter, proposed several decades ago in the context of high energy physics \cite{peccei-quinn, weinberg, wilczek}, and more recently for paired ground states with $p+is$ symmetry in various three-dimensional doped narrow gap semiconductors, such as Cu$_x$Bi$_2$Se$_3$, Sn$_{1-x}$In$_x$Te \cite{goswami-roy-axion} and and a parity and time-reversal odd Kondo singlet order in a strongly correlated topological Kondo insulators~\cite{roy-dzero}. However, experimental detection of axions has remained elusive thus far. In this paper, we have shown that axionic phase can be realized in various condensed matter systems at low-$T$, such as Dirac (with Zeeman coupling) and Weyl semimetals, but, for weak repulsive interactions, when these systems are placed in strong magnetic fields.

The complex axionic mass can be represented as $m(\mathbf{r})=|\Delta(\mathbf{r})|$ $\exp[-i \mathbf{Q} \cdot \mathbf{r} -i \phi(\mathbf{r})]$, where $\mathbf{Q}$ is the separation of Weyl nodes in the BZ. After a local chiral transformation on the fermion field $\Psi(\mathbf{r}) \to \Psi(\mathbf{r}) \exp[ i \left( \mathbf{Q} \cdot \mathbf{r} + \phi(\mathbf{r}) \right) \gamma_5/2]$ the complex mass becomes \emph{real}. However, the path integral measure and the action is not invariant under such chiral transformation~\cite{fujikawa} and results in an anomalous magneto-electric term 
\begin{eqnarray}
{\cal S}_{ax} &=& n \times \frac{e^2}{32 \pi^2}\int dt \; d\mathbf{r} \: \epsilon^{\mu \nu \rho \lambda} \left[ \mathbf{Q} \cdot \mathbf{r} + \phi(\mathbf{r}) \right] \: F_{\mu \nu} \; F_{\rho \lambda}, \nonumber \\
&=& n \times \frac{e^2}{4 \pi^2} \int dt \; d\mathbf{r} \left[ \mathbf{Q} \cdot \mathbf{r} + \phi(\mathbf{r}) \right] \mathbf{E} \cdot \mathbf{B},
\end{eqnarray}
where $F_{\mu \nu}$ is the electro-magnetic field strength tensor. The coefficient of the magneto-electric term $n=1,2,3$, respectively for WSM, double-WSM and triple-WSM.

When magnetic field is applied along $z$-direction, corresponding charge density is given by
\begin{equation}
j_0=\frac{n e^2}{4 \pi^2} \: \: \left( q_z + \partial_z \phi \right) \: B. 
\end{equation}
The first term gives rise to layer quantum Hall effect with thickness $2 n \pi/q_z$ accounting for the contribution from one-dimensional chiral surface states, residing at the boundary of WSMs, which can be observed in ARPES experiments. The second term is new and arises from contribution of scattered states bound to the bulk of the WSM, the ZLLs. Decomposing this term as $\frac{n e B}{2 \pi} \times \frac{e}{2 \pi} \partial_z \phi$, we find that second term is corresponds to one-dimensional charge-density, while the first one accounts for the degeneracy of ZLL in WSMs ($n=1$), double-WSMs ($n=2$) and triple-WSM ($n=3$). Effect of this term shows up only where $\partial_z \phi$ jumps, i.e., at the interface of WSMs with vacuum. Hence, the bulk anomalous term ${\cal S}_{ax}$ gives rise to surface Hall conductivity in WSMs, and through bulk-boundary correspondence theory remains anomaly free. Therefore, by comparing the separation of Weyl nodes from ARPES and surface Hall conductivity one can extract the contribution from the second term in ${\cal S}_{ax}$ arising from the bulk axionic CDW order.

So far, we have discussed the effect of axionic density-wave in the uniform phase. The $U(1)$ CDW order can also allow the existence of topological defects, e.g. \emph{line vortex} or \emph{axion-string}, along the $z$-direction. For simplicity we consider the vorticity to be \emph{one} and restrict ourselves in the \emph{dilute} vortex limit. The line vortex accommodates $n$-number of chiral one-dimensional dispersive gapless fermionic modes in its core that carries nondissipative electric current in $z$-direction, determined by one dimensional chiral anomaly 
\begin{equation}
j_z=n \times \frac{e^2 E_z}{2 \pi}, 
\end{equation}
with $n=1, 2,3$ respectively for WSM, double-WSM and triple-WSM \cite{roy-goswami-z2}. This current in turn is pumped from the bulk radially, which is captured by the bulk axionic term ${\cal S}_{ax}$, according to the \emph{Callan-Harvey} mechanism \cite{callan-hervey, magneticTI-2}. The exact solution of the dispersive modes can readily be obtained upon multiplying solutions of precise zero energy modes bound to a \emph{point vortex} in the $x-y$ plane \cite{lu-herbut}, with the plane-wave factor $\exp(i k_z z)$. Existence of such one-dimensional gapless dispersive modes will manifest in a T-linear specific heat in the ordered phase when it accommodates line-vortex, which is distinct from T$^3$ specific heat in the normal phase of three-dimensional WSMs.

\section{Summary and conclusions}

To summarize, we here propose that both Dirac and Weyl semimetals can undergo a weak coupling instability towards the formation of a CDW order in the presence of strong magnetic fields. Due to separation of Weyl nodes, which naturally occurs in WSM and due to the Zeeman coupling in trivial as well as topological DSM, the CDW order spontaneously breaks the translational symmetry and represents an axionic phase of matter. In this work, we demonstrate the effect of such mass generation on the renormalization of charge, diamagnetic susceptibility and also analyzed the scaling behavior of the spectral gap with the strength of sub-critical interactions and magnetic field. Similar mechanism has been argued to be operative in double- and triple-WSMs, where due to the additional degeneracy of the ZLL a larger gap can possibly be realized. Furthermore, we staunchly argued that among that charge- and spin-density waves, both of which can led to a spectral gap at the Weyl points, the former one wins energetically since it pushes the filled LLs down in energy. Thus our proposed axionic phase of matter can be realized in topological DSMs, such as Cd$_2$As$_3$~\cite{weylexperiment1}, Na$_3$Bi~\cite{weylexperiment2}, recently found WSMs in TaAs~\cite{taas-1,tasas-2,taas-3}, NbAs~\cite{nbas-1}, TaP~\cite{tap-1}, YbMnBi$_2$~\cite{borisenko}, Sr$_{1-y}$MnSb$_2$~\cite{chiorescu}, as well as in various three-dimensional strong-spin orbit coupled materials, such as Bi$_2$Se$_3$, when these systems are placed in close vicinity of the quantum critical point between topological and normal insulating phases. In addition, proposals for realizing double-WSM in HgCr$_2$Se$_4$ \cite{bernevig, Fang-HgCrSe, nagaosa}, SrSi$_2$~\cite{srsi2} give a genuine hope that axionic CDW orders, corrections to DMS, anomalous transport behavior can be observed in various members of the Weyl family in near future. 

Besides the uniform ground state, we also considered the role of topological defects in the ordered phase. Due to the associated $U(1)$ angle in the axionic-CDW phase, the ordered phase can support line vortex, also known as axion-string. In the dilute vortex limit with single vorticity, we show that such defects hosts $n$-number of one-dimensional gapless propagating mode, localized in its core, where $n=1,2,3$ for WSM, double-WSM and triple-WSM, respectively. Thus the number of gapless modes is intimately tied with the topological invariant of the system. Such dispersive mode carries non-dissipative electric current which in turn is supplied radially from the bulk, through the Callan-Harvey mechanism.    

As a final remark, we comment on the role of disorder. Notice that axionic-CDW in Weyl, double- and tripple-Weyl semimetals breaks the translational symmetry due to momentum space separation of the Weyl nodes. Most likely, periodicity of such CDW order is incommensurate with the lattice periodicity. Therefore, axionic-CDW can be susceptible to generic \emph{disorders}~\cite{imry}, which may reduce the ordering temperature considerably. Nevertheless, we expect that in sufficiently clean systems and strong magnetic fields, a sizable mass gap can be observed at low-$T$.

\acknowledgements

This work is supported by the start up grant of J. D. S. from the University of Maryland. B. R. is thankful to P. Goswami, P. Armitage and N. Nagaosa for many fruitful discussions. We thank V. Juri\v ci\' c for critical reading of the manuscript.

\onecolumngrid
\appendix

\section{Derivation of gap equation}

We devote this Appendix to present a detail derivation of the gap equation, quoted in Eq.~(\ref{gap-compact}). The condensation energy in the presence of CSB mass is given by 
\begin{equation}
E=\frac{\Delta^2}{4 g}-\frac{B}{2 \pi} \int^{\infty}_{-\infty} \frac{d k_z}{2 \pi} \left[\sqrt{\Delta^2+k^2_z} + 2 \sum_{n \geq 1}\sqrt{2 n B + k^2_z + \Delta^2} \right],
\end{equation}  
as shown in Eq.~(\ref{freeen-gap}) in the paper. Minimizing $E$ with respect to $\Delta$ we obtain the gap equation
\begin{eqnarray}
\frac{1}{g}&=&\frac{B}{\pi} \int^{\infty}_{-\infty}\frac{d k_z}{2 \pi} \left[ \frac{1}{\sqrt{\Delta^2+k^2_z}}+ \sum_{n \geq 1}\frac{2}{\sqrt{2 n B + \Delta^2+k^2_z}} \right]
=\frac{B}{\pi^{3/2}} \int^{\infty}_{\Lambda^{-2}} \frac{ds}{\sqrt{s}} \int^{\infty}_{-\infty}\frac{d k_z}{2 \pi} e^{-s (k^2_z + \Delta^2)} \left[1 + 2 \sum_{n \geq 1} e^{-s (2 n B)} \right] \nonumber \\
&=&\frac{B}{\pi^2} \int^{\infty}_{\Lambda^{-2}} \frac{ds}{s} \left[-\frac{1}{2}+ \sum_{n \geq 0} e^{-s (2 n B)} \right] e^{-s \Delta^2}
=\frac{B}{\pi^2} \int^{\infty}_{\Lambda^{-2}} \frac{ds}{s} \left[\frac{e^{2 s B}}{e^{2 s B}-1} -\frac{1}{2}\right]  e^{-s \Delta^2}
=\frac{B}{2 \pi^2} \int^{\infty}_{\Lambda^{-2}} \frac{ds}{s} e^{-s \Delta^2} \coth(sB). \nonumber \\
\end{eqnarray}
The above gap equation shows ultraviolet (UV) divergence as we take the UV cut-off $\Lambda \to \infty$. To regulate such divergence, after taking $2 \pi^2/g \to g$, we can rewrite the gap equation as 
\begin{eqnarray}
\frac{1}{g}-\Lambda^2 \int^{\infty}_{0} ds \frac{K(s)}{s^2}= -\int^{\infty}_{\Lambda^{-2}} \frac{ds}{s^2} \left[1- B s e^{-s \Delta^2} \coth(s B) \right],
\end{eqnarray}
where the function $K(s)$ satisfies the asymptotic properties $K(s \to 0)=0$ and $K(s \to \infty)=1$, otherwise arbitrary. In terms of dimensionless variables two $\Delta/(\Lambda v) \to m$, $B/\Lambda^2 \to B$, with $m,b \ll 1$, the above gap equation reduces to 
\begin{equation}
\delta+ I_1(m,B)+I_2 (m,B)=0. 
\end{equation}
For weak interactions $y=B/m^2 \gg 1$, and expanding the functions $I_1(m,B)$ and $I_2 (m,B)$ for small $m$ and $B$, as well as large $y$, we obtain 
\begin{eqnarray}
I_2(m,B) &=& -B \int^{\infty}_{1} \frac{ds}{s} e^{-s \; m^2}=B \left[ \gamma_E + 2 \log(m) - m^2 + \frac{m^4}{4} + {\cal O} (m^6)\right], \\
I_1(m,B) &=& 2 B\int^{\infty}_{2 B} \frac{ds}{s^2}\left[ 1-\frac{s e^{-s/(2 y)}}{e^t-1}\right]
=2B\bigg[ \int^{\infty}_0 \frac{ds}{s^2} \left(1- \frac{s+\frac{1}{2}s^2}{e^s-1} \right) + \frac{1}{2} \left( 1+ \frac{1}{y} \right) \int^{\infty}_{2B} ds \; \left( \frac{1}{e^s-1} \right) \nonumber \\
&-& \frac{1}{2} \cdot \; \frac{1}{(2 y)^2} \int^{\infty}_0 ds \left(\frac{s}{e^s-1} \right) +\frac{1}{6} \cdot \frac{1}{(2 y)^3} \int^{\infty}_0 ds \left( \frac{s^2}{e^s-1}\right) -{\cal O}(y^{-4}) \bigg] \nonumber \\
&=& 2B \bigg[ 0.63-\frac{0.21}{y^2} +\frac{0.05}{y^3} + \frac{1}{2} \left( 1+ \frac{1}{y}\right) \bigg\{ -\log(2B)+ B -\frac{B^2}{6}\bigg\} +{\cal O}(y^{-4},B^6) \bigg], \\
\end{eqnarray}
where
\begin{eqnarray}
\delta = \frac{1}{g \Lambda^2} -\int^{\infty}_{0} ds \frac{K(s)}{s^2} \equiv \frac{1}{g \Lambda^2}-\frac{1}{\Lambda^2 g_c}, \: \: \mbox{and} \: \: \frac{1}{g_c} = \Lambda^2 \int^{\infty}_{0} ds \frac{K(s)}{s^2}, \nonumber
\end{eqnarray}
is the zero magnetic field critical strength of the interaction for CSB ordering. Therefore, $\delta$ measures the deviation from the zero magnetic field critical point ($\delta=0$) and $\delta>0$ corresponds to subcritical interaction, i.e., $g<g_c$.

\twocolumngrid


\begin{thebibliography}{}
\bibitem{comment-periodicity} In a lattice, since $|Q_3| \leq \pi$, free energy needs to be minimized for $Q_3$ with an underlying lattice potential. Any finite $Q_3$ changes the location of the Weyl nodes, however, without qualitatively affecting our results.
\end{thebibliography}
\end{document}